\documentclass[aps, prb, showpacs, twocolumn, amsmath, letterpaper]{revtex4-2}

\usepackage{graphics}
\usepackage{graphicx}

\usepackage{amssymb}
\usepackage{bm}

\usepackage[english]{babel}

\usepackage{amsmath}
\usepackage{graphicx}
\usepackage{wrapfig}
\usepackage[colorlinks=true, allcolors=blue]{hyperref}

\begin{document}

\title{Revealing a 3D Fermi Surface pocket and Electron-Hole Tunneling in UTe$_{2}$ with Quantum Oscillations}

\author{Christopher Broyles$^{1}$, Zack Rehfuss$^{1}$, Hasan Siddiquee$^{1}$, Jiahui Althena Zhu$^{1}$, Kaiwen Zheng$^{1}$, Martin Nikolo$^{2}$, David Graf$^{3}$, John Singleton$^{4}$, Sheng Ran$^{1}$}

\affiliation{$^1$ Department of Physics, Washington University in St. Louis, St. Louis, MO 63130, USA
\\$^2$ Department of Physics, Saint Louis University, St. Louis, MO, 63103, USA
\\$^3$ National High Magnetic Field Laboratory, Florida State University, Tallahassee, FL 32310, USA
\\$^4$National High Magnetic Field Laboratory, Pulse Field Facility, Los Alamos National Laboratory,
 Los Alamos, New Mexico, 87545, USA}

\begin{abstract}

Spin triplet superconductor UTe$_{2}$ is widely believed to host a quasi-two-dimensional Fermi surface, revealed by first-principles calculations, photoemission and quantum oscillation measurements. An outstanding question still remains as to the existence of a three-dimensional Fermi surface pocket, which is crucial for our understanding of the exotic superconducting and topological properties of UTe$_{2}$. This 3D Fermi surface pocket appears in various theoretical models with different physics origins, but has not been unambiguously detected in experiment. Here for the first time we provide concrete evidence for a relatively isotropic, small Fermi surface pocket of UTe$_{2}$ via quantum oscillation measurements. In addition, we observed high frequency quantum oscillations corresponding to electron-hole tunneling between adjacent electron and hole pockets. The coexistence of 2D and 3D Fermi surface pockets, as well as the breakdown orbits, provides a test bed for theoretical models and aid the realization of a unified understanding of superconducting state of UTe$_{2}$ from the first-principles approach.  


\end{abstract}

\maketitle{}

Spin triplet superconductivity offers a promising route to topologically protected quantum computing, with Majorana Fermions as surface state excitations~\cite{Sato2017,Kallin2016}. The recent discovery of spin triplet superconductivity in UTe$_{2}$ has prompted intense research efforts~\cite{ran2019nearly,ran2019extreme}. Spin triplet pairing is strongly suggested by the extremely large, anisotropic upper critical field $H_{\textup{c2}}$~\cite{ran2019nearly} and the temperature independent NMR Knight shift in the superconducting state~\cite{nakamine2019superconducting,PhysRevB.103.L100503,fujibayashi2022superconducting}. Time reversal symmetry breaking is indicated by the observation of finite Kerr rotation ~\cite{hayes2021multicomponent} as well as the chiral in-gap bound states shown by scanning tunneling spectroscopy~\cite{jiao2020chiral}. Charge density waves have been observed in the normal state of UTe$_{2}$ from scanning tunneling spectroscopy measurements~\cite{aishwarya2022magnetic}, which develop into an unprecedented spin triplet pair density wave in the superconducting state\cite{gu2022detection}.

\begin{figure*}[ht]
    \includegraphics[width=1\linewidth]{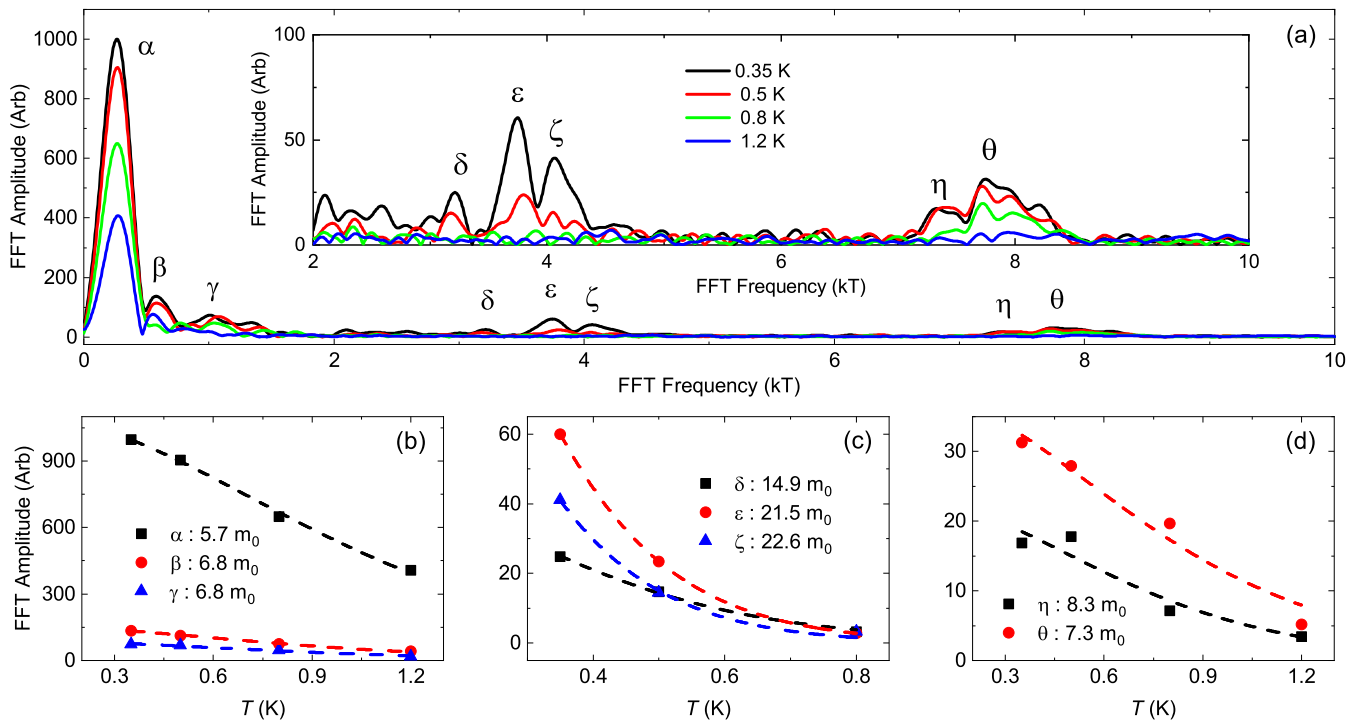}
    \caption{\textbf{(a)} The FFT spectrum of the background subtracted signal, $\Delta f_{TDO}$, is plotted for multiple temperatures, with the magnetic field oriented close to the $c$-axis. The inset shows the mid and high frequency peaks with better resolution.~\textbf{(b-d)} The temperature dependent dampening of the FFT peaks are fitted with Lifshitz–Kosevich formalism  to calculate effective mass $m^*$.}
\end{figure*} 

A knowledge of electronic band structure is essential in the understanding of superconducting and topological properties of UTe$_{2}$. Despite extensive efforts~\cite{xu2019quasi,ishizuka2019insulator,PhysRevB.103.104504,PhysRevB.103.094504,kang2022orbital,choi2022correlated}, a unified picture has not been achieved. A quasi-2D Fermi surface, oriented along $k_x$ and $k_y$ , was revealed by first-principles calculations~\cite{xu2019quasi,ishizuka2019insulator,PhysRevB.103.104504}, and supported by angle-resolved photoemission (ARPES) and de Haas van Alphen (dHvA) oscillation measurements~\cite{miao2020low,aoki2022first}. On the other hand, this large 2D Fermi surface alone does not explain various properties of UTe$_{2}$, e.g., the nontrivial topology of the superconductivity~\cite{xu2019quasi,ishizuka2019insulator}, the multiple superconducting states~\cite{Ran2020,Thomas2020,braithwaite2019multiple}, the Fermi surface reconstruction at the metamagentic transition~\cite{Niu2020a, Niu2020}, and the nearly isotropic electronic transport~\cite{PhysRevB.106.L060505}. This spurred searches for additional Fermi-surface sections, likely with a 3D character. A reexamination of the Kondo interaction within DFT+DMFT calculations predicted a 3D isotropic Fermi-surface pocket surrounding the $\Gamma$-point~\cite{choi2022correlated}. The electron and hole Fermi-surfaces are also reconstructed to enclose the corners of the Brillouin zone~\cite{choi2022correlated}. In contrast, a tight binding model approach preserves the quasi-2D structure calculated from DFT+$U$ while introducing a new Fermi surface with 3D character at the $k_z = \pi/2c$ plane~\cite{PhysRevB.103.094504}. 3D Fermi-surface sections may have been observed in ARPES measurements~\cite{miao2020low,fujimori2019electronic}. However, the noise level of the data leaves the conclusions ambiguous, e.g., a close examination reveals that the soft x-ray ARPES data~\cite{fujimori2019electronic} could as well match predictions of quasi-2D Fermi surface sections.

Here we report quantum oscillations of UTe$_{2}$ observed on high quality single crystals grown using molten salt flux method~\cite{sakai2022single}. Most importantly, we provide the first concrete evidence for a relatively isotropic, small Fermi surface pocket, in addition to the quasi-2D cylindrical Fermi surface. We also observed quantum oscillations with very high frequencies, consistent with breakdown orbits due to the tunneling between electron and hole pockets. The 3D Fermi surface holds the promise for the understanding of many exotic properties of UTe$_{2}$, such as the topology of the superconducting state~\cite{choi2022correlated}, the nature of magnetic fluctuations~\cite{PhysRevB.103.094504}, and the mechanism of the high field induced superconductivity. Our observations apply strong constrains on the theoretical understanding of UTe$_{2}$. Any successful model of UTe$_{2}$ needs to account for all three salient features: the 2D cylindrical Fermi-surface section, small isotropic Fermi-surface sections and adjacent pockets allowing for tunneling.

The quantum oscillations of UTe$_2$ were measured using an inductance coil as part of a tunnel diode oscillator (TDO) circuit, where the resonant frequency, $f_{TDO}$, is coupled to the sample magnetization and conductivity (see SI for details~\cite{suppinfo}). Figure~1a plots the fast Fourier transform (FFT) of $\Delta f_{\rm TDO}$ with magnetic field oriented close to the $c$ axis. There are three groups of well resolved frequencies, below 1~kT, around 4~kT and 8~kT. As expected for magnetic quantum oscillations, the amplitudes of all of these frequencies decrease with increasing temperature, in agreement with the expectations of Fermi-Dirac statistics. The effective masses are calculated from the temperature dependent dampening using the Lifshitz–Kosevich (L-K) formalism~\cite{DShoenberg_1988}. The obtained values are large for all the frequencies, in the range of $5-25~m_0$, where $m_0$ is the mass of the free electron, consistent with heavy electronic states. Note that since the fitting is based on only three or four data points, the error associated with the effective mass is expected to be relatively large.



The mid-range frequencies around 4~kT are in a good agreement with previous dHvA oscillations seen in magnetic fields of up to 14~T~\cite{aoki2022first}. These frequencies are consistent with those of the GGA+$U$ calculations with $U$ = 2~eV, which predicts quasi-2D electron and hole Fermi surface with a cylindrical shape~\cite{ishizuka2019insulator}. Based on these calculations, F$_{\delta}$ = 3.2~kT corresponds to a hole pocket oriented along $k_x$, and F$_{\epsilon}$ = 3.7~kT and F$_{\zeta}$ = 4.1~kT to an electron pocket oriented along $k_y$~\cite{ishizuka2019insulator, xu2019quasi}. 

Further knowledge is gained on the topology of the Fermi surface from the angle dependence of the quantum oscillation frequencies. Here, two samples are measured: sample S1, which rotates from $H \parallel c$ towards the $b$-axis, and sample S2, which rotates from $H \parallel c$ towards the $a$-axis. To further confirm the rotation angle, the critical field for the metamagnetic transition was used for sample S1 and the superconducting critical field was used for sample S2, as shown in Fig.~2d. The FFT peak positions for both samples are collected in Fig.~2c, with the $c-a$ axis rotation plotted against recent dHvA oscillations~\cite{aoki2022first}. The $c-a$ axis rotation provides good agreement, with the both electron and hole orbits showing cylindrical angle dependence. Similar cylindrical angle dependence is observed for the $c-b$-axis rotation in the low angle range, while the FFT peaks quickly drop to the noise level for angles above 30$^{\circ}$.  




Our key finding is the observation of quantum oscillations with new frequencies in two frequency ranges, below 1~kT and around 8~kT. First we investigate the frequencies around 8~kT. Figure~2a shows 2~kT high pass filtered $\Delta f_{TDO}$ from sample S1, with a 5$^{\circ}$ rotation from $c$ towards the $b$-axis ($\theta_b = 5$). Below 36~T, there is a prevalent 7.7~kT oscillation, which is shrouded by a larger magnitude 4~kT oscillation as $H$ increases. The beating phenomenon in the high field range is due to a collection of frequencies around 4~kT. The FFT shows at least two well resolved frequencies around 8~kT. The peak position is tracked until $\theta_b = 20$, depicting a similar angle dependence to its mid-frequency counterparts.


\begin{figure}[t]
    \includegraphics[width=1\linewidth]{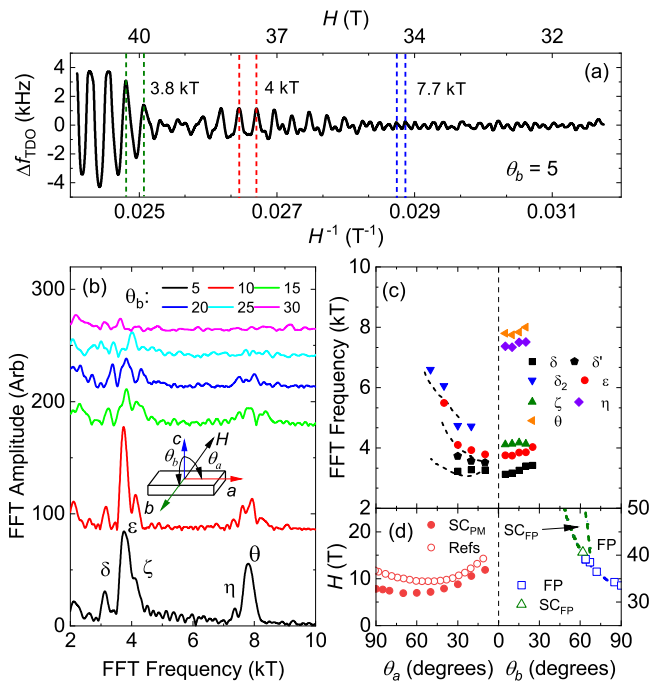}
    \caption{\textbf{(a)} The background subtracted signal, $\Delta f_{TDO}$, with a high pass filter of 2~kT, is plotted as a function of $H^{-1}$, with magnetic field oriented 5$^{\circ}$ from the $c$ axis toward $b$ axis. There are three distinct oscillations, which correspond to F$_{\epsilon}$ (green), F$_{\zeta}$ (red), and F$_{\theta}$ (blue). \textbf{(b)} Multiple spectra are plotted as the magnetic field is rotated from the $c$ axis towards $b$ axis, indicated by $\theta_b$. \textbf{(c)} The peak position of the mid and high frequencies is plotted as the magnetic field is rotated from the $c$ axis towards $a$ axis (left), and from the $c$ axis towards $b$ axis (right). The $c-a$ rotation is compared with literature~\cite{aoki2022first}, plotted with dashed lines. \textbf{(d)} The $c-a$ rotation is traced with $H_{c2}$ of the low field superconducting phase at 350~mK (solid), compared with measurements at 70~mKfrom literature (open)~\cite{aoki2022first}. The $c-b$ rotation is traced with the critical field of the metamagnetic transition $H_{FP}$, along with the phase boundaries from the literature (dashed lines)~\cite{ran2019extreme}.  }
\end{figure}

A frequency of 7.7~kT would correspond to a Fermi surface area of ${\rm 0.74~\AA^{-2}}$, which is $48\%$ of the Brillouin zone in the $k_x$-$k_y$ plane. There is no experimental nor theoretical indication of such a large Fermi-surface section ~\cite{fujimori2019electronic,miao2020low,xu2019quasi,ishizuka2019insulator}, coexisting with the electron and hole pockets discussed above. There are a few other mechanisms that can potentially give rise to high oscillation frequencies: high order harmonics, magnetic breakdown, and quantum interference~\cite{shoenberg2009magnetic}. The high-frequency peaks, F$_{\eta}$ and F$_{\theta}$, are close to being second harmonics of the mid-frequency peaks. However, the temperature dependence of the $n$th harmonic should give an apparent effective mass that is $n$ times the mass of the fundamental $(n=1)$ frequency~\cite{DShoenberg_1988,shoenberg2009magnetic}. The apparent effective masses for the high frequencies are smaller than those of the 4~kT frequencies, ruling out this possibility.


In high magnetic fields, electrons could tunnel between adjacent pockets, giving rise to new quantum oscillations in breakdown orbits~\cite{shoenberg2009magnetic}. The breakdown field $B_c \propto \Delta k^2$, with $\Delta k$ being the gap between adjacent pockets of the Fermi surface, is typically too large for this effect to be observed in most materials. In UTe$_{2}$, adjacent pockets are close to each other in $k$-space, making it possible to observe new oscillations in relatively low magnetic field (starting from around 30~T). Our analysis indicates that the each of the high frequencies is roughly the addition of two frequencies in the 4~kT range, consistent with the breakdown orbits. There is an additional increase ($\sim$ 400~T) in FFT frequency corresponding to the tunneling length in $k$-space. Note that in case of UTe$_{2}$ the tunneling happens between adjacent electron and hole pockets~\cite{Delft2018, Brien2016}, in which electrons move in opposite directions.    


In Fig. 3 we present two scenarios for the tunneling. In the first case, figure of eight pattern, electrons keep the same direction when tunneling between electron and hole pockets. This tunneling is allowed by semiclassical theory. The resulting oscillation frequency of the breakdown orbit equals the difference of the involved frequencies $f_a - f_b$ (due to the opposite enclosed areas defined by electron and hole paths, detailed in SI) and the effective mass equals the sum of the involved effective masses $m_a + m_b$. This is opposite to what we observed. As a matter of fact, due to the large resulting effective mass, it requires much lower temperatures to observe frequencies due to this type of tunneling. The frequency we observed, $f = f_a + f_b$, seems to correspond to the second case, known as Stark quantum interference, which has been observed in other materials~\cite{Sasaki1990,Harrison1996,Delft2018}.
This effect is due to interference between alternative quasiparticle orbits around different Fermi-surface sections that are weakly connected via tunneling~\cite{Stark1977,Morrison1981}. The interference leads to an apparent effective mass that is the difference between the effective masses of the constituent orbits, $m = m_a - m_b$, in agreement with our observations (see SI~\cite{suppinfo}). 

\begin{figure}[b]
    \includegraphics[width=1\linewidth]{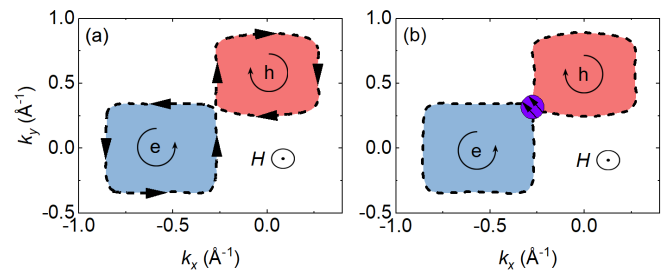}
    \caption{A cartoon diagram of the breakdown orbits due to the electron-hole tunneling between the adjacent pockets of 2D Fermi surface of UTe$_{2}$, taken from reference~\cite{ishizuka2019insulator}. The orbit depicted in \textbf{(a)} is allowed by semi-classical theory and results in the subtraction of frequency and sum of effective mass, $f = f_a - f_b$ and $m = m_a + m_b$. \textbf{(b)} Stark quantum interference between the orbits about the electron and hole pockets results in the sum of frequency and subtraction of effective mass, $f = f_a + f_b$ and $m = m_a - m_b$, consistent with our observation.}
\end{figure}


Now we turn to the other group of new frequencies. In the low frequency regime, we observed oscillations with $F_{\alpha} = 270$~T, $F_{\beta} = 680$~T, and $F_{\gamma} = 1000$~T. The signature of the lowest frequency oscillation, $F_{\alpha}$, is prominent for both samples S1 and S2. Note that the low frequency oscillations are readily visible even in the raw data of S2, displayed in Fig. 4a. Similar oscillations for S1 are plotted in Fig.~4b and 4c, with field oriented along the $c$- and $b$-axis respectively. The low frequency peaks could as well come from magnetic breakdown of 4~kT orbits in the form of figure of eight. However, the effective mass would be $m_a + m_b$, much higher than the observed values. Therefore the low frequency oscillations must pertain to a real carrier orbit on the Fermi surface.


 

\begin{figure}[t]
    \includegraphics[width=1\linewidth]{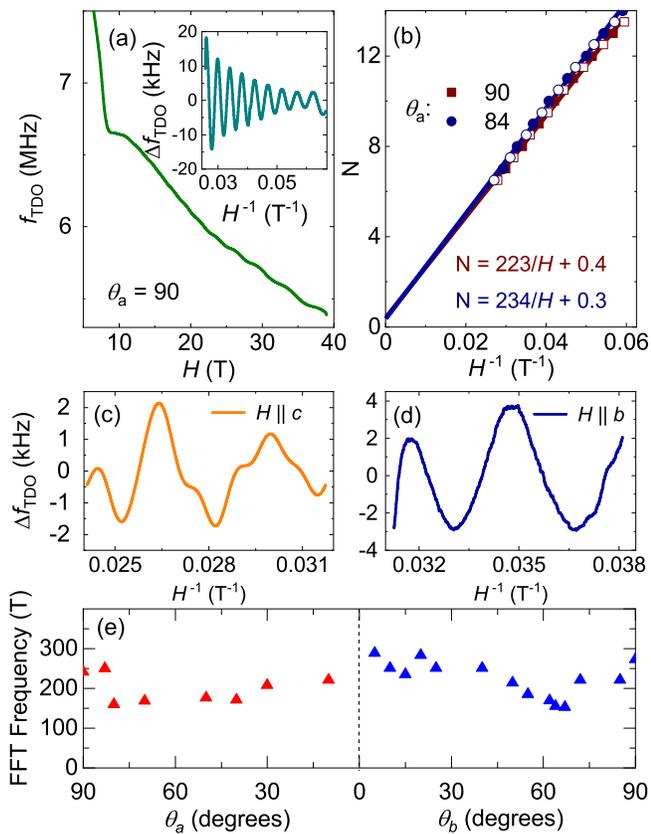}
    \caption{\textbf{(a)} The raw data of TDO measurements for sample S2, with magnetic field oriented along the $a$-axis. The inset displays the background subtracted signal. \textbf{(b)} The Landau level fan diagram is constructed for two angles near the $a$-axis by assigning $n$ to maximums and $n+1/2$ to minimums of the oscillations. The phase shift, $\Phi$, is calculated by extrapolating the linear fit to 0 $H^{-1}$. The background subtracted data, $\Delta f_{TDO}$, with a low pass filter of 2~kT, for sample S1 is plotted for magnetic field oriented along ~\textbf{(c)} the $c$-axis and~\textbf{(d)} along the $b$-axis. \textbf{(e)}The peak position of the low frequencies is plotted as the magnetic field is rotated from the $c$ axis towards $a$ axis (left), and from the $c$ axis towards $b$ axis (right).}
\end{figure} 


The geometry of this small Fermi-surface pocket is investigated using the field-angle dependence in Fig.~4e. There is a small dip in frequency at $\theta_b = 60^{\circ}$, which is consistent for multiple field sweeps. This angle range coincides with the field induced superconducting state ~\cite{ran2019extreme}. Other than that, $F_{\alpha}$ does not have significant angle dependence, indicating a relatively isotropic Fermi-surface section. Note that for a large angle range, only a few cycles of oscillations were observed, therefore the frequency determined from FFT may not be accurate. For comparison, we also use sinusoidal fit to obtain the frequency, which shows similar angle dependence of $F_{\alpha}$ (see SI). A sufficient number of cycles were observed near the $a$-axis, which allows to construct Landau level fan diagrams for two angles, $\theta_a$ = 90$^{\circ}$ and 84$^{\circ}$, shown in Fig. 4b. The linear extrapolation for both angles show a non-zero y-intercept, indicating a possible topologically nontrivial orbit.  


Our observation of low frequency quantum oscillations does not contradict other quantum oscillation studies on UTe$_{2}$~\cite{aoki2022first,aoki2023haas,eaton2023quasi2d}, which have only reported frequencies above 3~kT. In the dHvA study via field modulation technique, there is a clear low frequency peak in the FFT spectrum presented in the supplemental information and in their follow-up study, with a temperature dependence consistent with quantum oscillations~\cite{aoki2022first,aoki2023haas}. However, the possibility for a small Fermi-surface section was overlooked. The other study that did not observe low frequency oscillations is based on torque measurements~\cite{eaton2023quasi2d}. It is well known that the oscillatory torque signal is strong for anisotropic Fermi surfaces, but very weak when the Fermi surface is almost isotropic. Therefore the absence of low frequencies in torque measurements is consistent with our observation of relatively isotropic Fermi-surface section. The TDO technique that we utilized in the current study is sensitive to the electric conductivity. The relatively isotropic Fermi-surface pockets that we observe have much lighter effective masses, and will therefore tend to dominate the electric transport properties, even though they are small. This also naturally reconciles the the quite isotropic electrical conductivity of UTe$_{2}$~\cite{PhysRevB.106.L060505} with the large quasi-2D Fermi-surface sections.

This relatively isotropic, small Fermi-surface section provides a test bed for different theoretical models of UTe$_{2}$. First-principles calculations have predicted a wide range of underlying band structures, largely due to the modeling complexity associated with $f$-electron strong correlations and Kondo lattice physics. Results of DFT+$U$ and GGA+$U$ calculations are very sensitive to the choice of $U$~\cite{ishizuka2019insulator}. A quasi-2D Fermi surface is predicted by GGA+$U$ for $U$ = 2~eV (DFT+$U$ for $U$ = 6~eV ), which is in a fairly good agreement with ARPES observation of rectangular pockets and dHvA oscillations of 4~kT frequency~\cite{miao2020low,aoki2022first}. When $U$ is decreased to 1.0~eV, GGA+$U$ calculations predict a tiny electron pocket around the $X$ point and a tiny hole pocket around the $R$ point. However, the large cylindrical Fermi-surface section no longer exists. Recent DFT + DMFT calculations predict a Kondo interaction induced Fermi surface reconstruction~\cite{choi2022correlated}. At low temperatures, the original quasi-2D Fermi surface morphs into another quasi-2D Fermi-surface section and a 3D small Fermi pocket closing the $\Gamma$ point. The renormalized 2D Fermi-surface section no longer has pockets close in the $k$-space, leaving the electron-hole tunneling impossible. It would require additional mechanism to explain the frequencies we observed around 8~kT. 


A promising route for 3D Fermi-surface section might be the tight binding model~\cite{PhysRevB.103.094504}, based on the results of DFT+$U$ calculations with an intermediate $U$. The 2D cylindrical Fermi-surface section is preserved in this model. In the meantime, due to the contribution of itinerant $f$-electrons, the hole Fermi-surface section is bent and encloses the $X$-point. It is conceivable that when hybridization between the Te 5$p$-orbital of the hole pocket and the U 5$f$-orbital is turned on, a gap will open and isolated small pockets will form around $X$-point that are almost spherical. In this scenario, all three groups of frequencies we observed could be captured in one model. Further theoretical investigation is required to verify this hypothesis. 



The 3D Fermi-surface pocket is crucial for the understanding of the exotic properties of UTe$_{2}$. Both GGA+$U$ and DFT+$U$ calculations predict a topologically trivial superconducting state for the 2D cylindrical Fermi surface~\cite{xu2019quasi,ishizuka2019insulator}. The addition of a 3D Fermi surface changes the occupation number at time-reversal invariant momenta, which in turn changes the winding number and could potentially lead to topological superconductivity~\cite{Sato2010,Fu2010}. In addition, new 3D Fermi surface might be responsible for the magnetic fluctuations and multiple order parameters observed in UTe$_{2}$. In the tight binding model, the Fermi surface at $k_z$ plane enhances ferromagnetic fluctuations, rather than antiferromagnetic ones, and extends the odd-parity symmetry from $B_{iu}$ to include the $A_u$ point-group, which transforms to the even-parity $A_g$ state under pressure~\cite{PhysRevB.103.094504}. This makes the small Fermi-surface sections key to the understanding of multiple superconducting states under pressure and the overall $P-T$ phase diagram~\cite{Ran2020,Thomas2020,braithwaite2019multiple,Lin2020,ran2021expansion,ambika2022possible}. Another open question about UTe$_{2}$ is the origin of the high-field superconducting phase existing inside the field-polarized state. Upon the metamagnetic transition into the field polarized state, a Fermi surface reconstruction has been suggested by the thermopower experiments~\cite{Niu2020a, Niu2020}, which likely involves a small, rather than large, Fermi-surface section. It is crucial to investigate how the small Fermi-surface pocket we observed here changes upon the metamagnetic transition, which is an undergoing project.

To summarize, we provide concrete evidence for a small 3D Fermi-surface section in UTe$_{2}$ via quantum oscillations measurements. The 3D small Fermi-surface pocket is crucial for our understanding of various exotic properties of UTe$_{2}$: nontrivial topological superconductivity, multiple order parameters, and the extremely high-field-induced superconducting state. Together with the observation of electron-hole tunneling, this discovery applies strong constrains to theoretical models, aiding the eventual realization of a unified understanding of the superconducting states of UTe$_{2}$. 




We would like to thank Andrew Wray, Yifeng Yang, Daniel Agterberg, Carlo Beenakker and Li Yang for fruitful discussions. Research at the National High Magnetic Field Laboratory NHMFL was supported by NSF Cooperative Agreement No.~DMR-1644779 and the State of Florida. 
JS thanks the DOE BED FWP {\it Science of 100~T} for support.

%

\end{document}